\documentclass[twocolumn,aps,prl,showpacs,amsmath,amssymb,floatfix]{revtex4-1}
\usepackage{graphicx}
\usepackage{dcolumn}
\usepackage{bm}
\usepackage{xcolor,tabu}
\usepackage[colorlinks=true,
            linkcolor=blue,
            urlcolor=blue,
            citecolor=blue]{hyperref}

\begin{document}

\title{Star-shaped Oscillations of Leidenfrost Drops}

\author{Xiaolei Ma}
\email{xiaolei.ma@emory.edu}
\author{Juan-Jos\'{e} Li\'{e}tor-Santos}
\author{Justin C. Burton}

\affiliation{Department of Physics, Emory University, Atlanta, Georgia 30322, USA}

\date{\today}

\begin{abstract} 
We experimentally investigate the self-sustained, star-shaped oscillations of Leidenfrost drops. The drops levitate on a cushion of evaporated vapor over a heated, curved surface. We observe modes with $n = 2-13$ lobes around the drop periphery. We find that the wavelength of the oscillations depends only on the capillary length of the liquid, and is independent of the drop radius and substrate temperature. However, the number of observed modes depends sensitively on the liquid viscosity. The dominant frequency of pressure variations in the vapor layer is approximately twice the drop oscillation frequency, consistent with a parametric forcing mechanism. Our results show that the star-shaped oscillations are driven by capillary waves of a characteristic wavelength beneath the drop, and that the waves are generated by a large shear stress at the liquid-vapor interface. 
\end{abstract}


\maketitle

The Leidenfrost effect can be easily observed by placing a millimeter-scale water drop onto a sufficiently hot pan. The drop will levitate on a thermally-insulating vapor layer and survive for minutes \cite{leidenfrost1756aquae,biance2003leidenfrost,gottfried1966film,quere2013leidenfrost}. For small drops, the geometry and dynamics of the vapor layer have been recently characterized \cite{burton2012geometry,caswell2014dynamics}. The complex interactions between the liquid, vapor, and solid interfaces have led to a broad range of applications such as turbulent drag-reduction \cite{vakarelski2011drag}, self-propulsion of drops on ratcheted surfaces \cite{linke2006self,lagubeau2011leidenfrost}, green nanofabrication \cite{abdelaziz2013green}, fuel combustion \cite{kadota2007microexplosion}, and thermal control of nuclear reactors \cite{van1992physics}. 

\begin{figure}[!]
\begin{center}
\includegraphics[width=3.4 in]{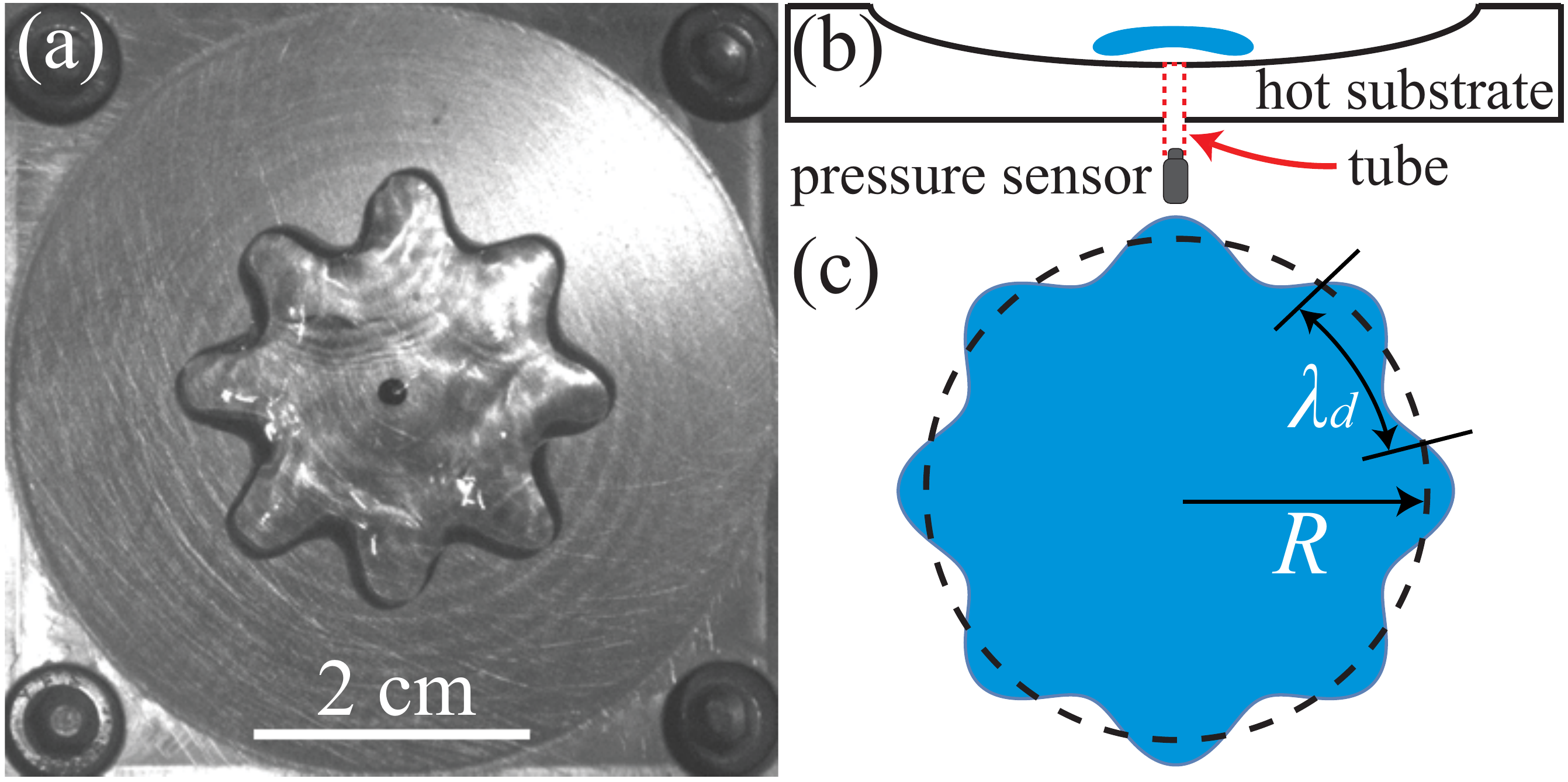}
\caption[]{(a) Top-down image of a star-shaped Leidenfrost water drop ($l_c=2.5$ mm, $R=1.4$ cm, $T_s=623$ K). The small, dark circle in the center of the substrate is used for pressure measurements. (b) Cross-sectional view of the experimental setup. (c) Top-down schematic of a star-shaped drop indicating the radius $R$ and wavelength $\lambda_d=2\pi R/n$. 
\label{setup}} 
\end{center} 
\end{figure}

Large Leidenfrost drops are well-known to form self-sustained, star-shaped oscillations (Fig.\ \ref{setup}a). Since the 1950's, a number of studies have investigated these oscillations, often with different conclusions as to their physical origin based on the complicated interplay between thermal and hydrodynamic effects in both the liquid and gas phases \cite{holter1952vibrations,strier2000nitrogen,tokugawa1994mechanism,adachi1984vibration,takaki1985vibration,celestini2014two,snezhko2008pulsating}. A simple underlying mechanism for the onset of star oscillations remains unknown. Drops subjected to external, periodic excitations can form star oscillations with a frequency half that of the external excitation due to a parametric coupling mechanism \cite{brunet2011star,shen2010parametrically}. However, if a parametric mechanism causes Leidenfrost stars, then the source of the periodic excitation is unclear. Recently, Bouwhuis \textit{et al.} investigated the star oscillations of drops levitated by an air flow over a porous surface \cite{bouwhuis2013oscillating}. They showed that the onset of star oscillations occurs when the flow rate of air beneath the drop reaches a threshold, suggesting that a hydrodynamic coupling between the gas flow and liquid interface initiates the oscillations.
       
Here we report measurements of star-shaped oscillations of six different liquids on a hot, curved surface. We observe stars with $n = 2-13$ lobes around the drop periphery. Although the number of observed modes depends on the liquid viscosity and substrate temperature, we find that the wavelength and frequency of the modes only depend on the capillary length, $l_c=\sqrt{\gamma/\rho_l g}$, where $\gamma$ and $\rho_l$ are the surface tension and density of the liquid, and $g$ is the acceleration due to gravity. The pressure near the center of the vapor layer oscillates at approximately twice the frequency of the drop oscillation, consistent with a parametric coupling mechanism. We suggest that this pressure variation stems from capillary waves of a characteristic wavelength, $\lambda_c\approx 3l_c$, generated by a large shear stress at the liquid-vapor interface beneath the drop.

In the experiment, most substrates were constructed from blocks of engineering 6061 aluminum alloy with dimensions 7.6 cm $\times$ 7.6 cm $\times$ 2.5 cm. The substrate temperature was controlled by resistive heaters embedded in the material. Six different liquids were used as Leidenfrost drops: deionized water, liquid N$_2$, acetone, methanol, ethanol and isopropanol. The physical properties of each liquid at the boiling point $T_ {b}$ are listed in Table \ref{tab1}. For water, the substrate temperature $T_s$ varied from 493 K to 773 K, while for ethanol, methanol, acetone, and isopropanol, $T_s$ was set to 523 K. The substrate for liquid N$_2$ was not heated due to its extremely low $T_ {b}$.

The upper surfaces of the substrates were machined into a spherical bowl-shape in order to suppress the buoyancy-driven Rayleigh-Taylor instability in the vapor layer and keep the drops stationary \cite{quere2013leidenfrost,trinh2014curvature,snoeijer2009maximum}. A cross-sectional view of the curved substrates is shown in Fig.\ \ref{setup}b. The curvature of each bowl-shaped surface was designed to satisfy $l_{c}/R_s$ = 0.03, where $R_s$ is the radius of curvature of the surface. For some experiments, a plano-concave, fused silica lens (focal length = 250 mm) was used as the heated substrate in order to allow for optical imaging of the vapor layer beneath the oscillating drop. 

We used a high-speed digital camera (Phantom V7.11, Vision Research) with a resolution of 132 pixels/cm to image the drops at 1000 frames/s. Recorded videos were then analyzed to obtain the frequency and wavelength of the star-shaped oscillations. An image of a water drop (mode $n=8$) is shown in Fig.\ \ref{setup}a, and a schematic of a typical star-shaped oscillation mode is shown in Fig.\ \ref{setup}c, indicating the radius, $R$, and wavelength of the standing wave, $\lambda_d$. In order to detect pressure variations beneath the drop, a pressure sensor (GEMS Sensors, response time: 5 ms, sensitivity: 2 mV/Pa) was connected to a small hole of diameter 1 mm at the center of the curved aluminum (Fig.\ \ref{setup}a). The pressure was sampled at 500-1000 Hz in order to accurately capture waveforms near the response time of the sensor. 

\begin{figure}[!]
\begin{center}
\includegraphics[width=3.4 in]{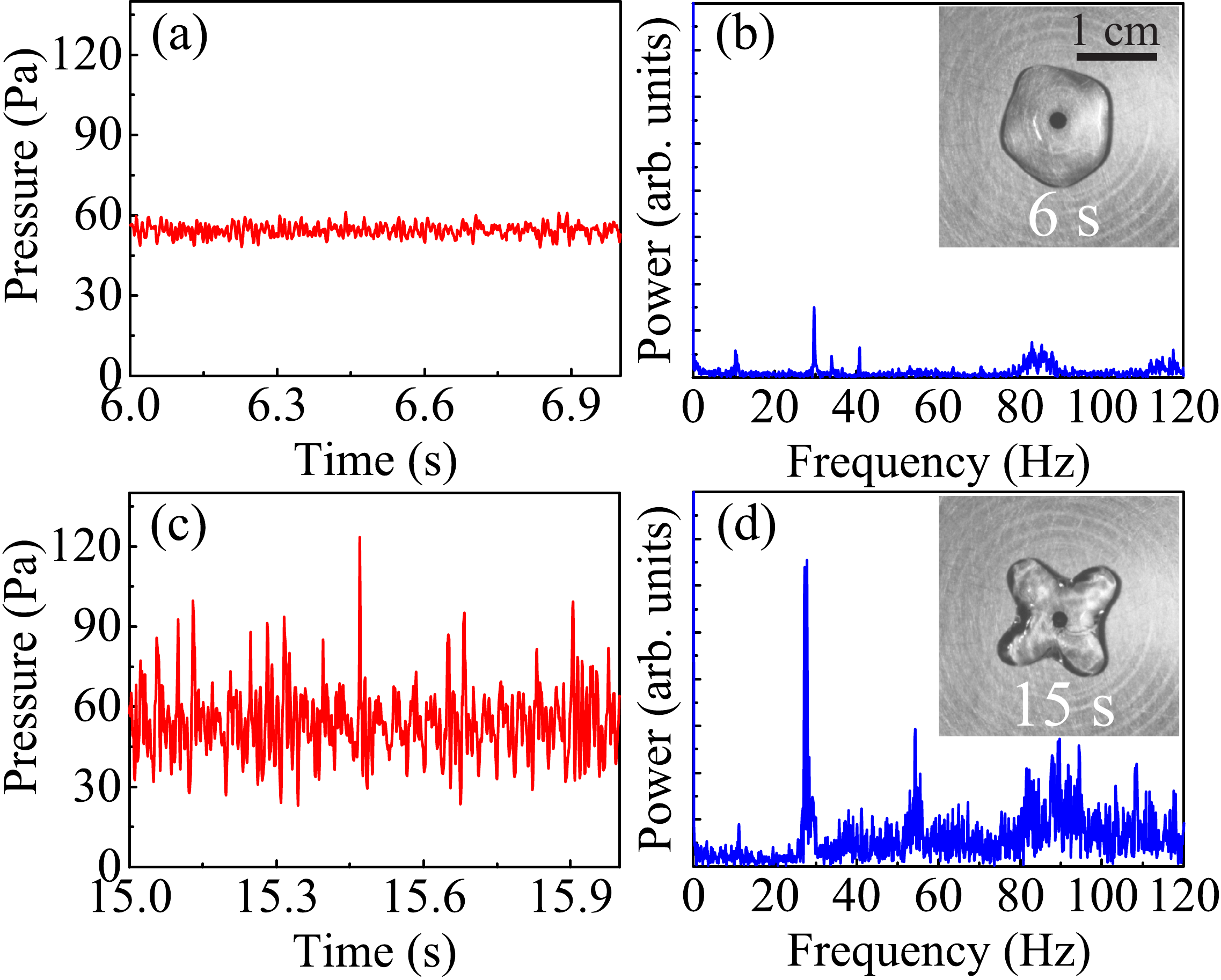}
\caption[]{(a) Pressure variations in the vapor layer beneath a Leidenfrost water drop ($T_s$ = 623 K) just before the initiation of the star-shaped oscillation. (b) Power spectrum of the pressure fluctuations over the time interval of $0-9$ s. (c) Pressure variations during a fully-developed $n=4$ mode. (d) Power spectrum of the pressure fluctuations over the time interval of $9-18$ s showing a sharp peak at $f_p\approx 28$ Hz. Both power spectra have the same linear vertical scale. The insets in (b) and (d) show snapshots of the drop shape, and the scale bar 1 cm applies to both images. 
\label{pressure_oscillation}
} 
\end{center} 
\end{figure}

For each observed star-shaped oscillation mode, we measured the azimuthal wavelength, $\lambda_d$, drop oscillation frequency, $f_d$, and the characteristic frequency of pressure oscillations, $f_p$. Figure \ref{pressure_oscillation}a shows the excess pressure (above atmosphere) before the initiation of a star-shaped oscillation for a water drop. For a very large drop, i.e., $R\gg l_c$, the drop can be approximated by a cylinder whose thickness, $h\approx2l_c$, is determined by a balance of surface tension and gravitational forces \cite{biance2003leidenfrost}. Thus the mean pressure required to levitate a large water Leidenfrost drop at the boiling point should be $\rho_l g h\approx47$ Pa. One can see that the pressure measured in the center is slightly larger than 47 Pa. This is expected since a radial pressure gradient is required to drive the viscous vapor from the center of the drop to the edge. 

\begin{table}[]
\caption{Physical properties of different liquids at the boiling point $T_b$ (K). Units are as follows: $\gamma$ (mN/m), $\rho_l$ (kg/m$^3$), $\eta_l$ (mPa $\cdot$ s), $l_c$ (mm). Data was taken from Ref. \cite{lemmon2011nist}.}
\centering
\begin{tabular*}{\columnwidth}{c @{\extracolsep{\fill}} cccccccc}
     \hline
     liquid                 &$T_{b}$ &$\gamma$      &$\rho_l$   &$\eta_l$   &$\l_{c}$  &Mode    &$Re_l$\\ 
           
     \hline
     water                 &373       &59.0	              &958	    &0.282	&2.5   &2-13          &1340\\
     liquid N$_2$	    &77         &8.90     	       &807	    &0.162	&1.1   &3-5,7        &539\\
     acetone	          &329       &18.2	       &727	    &0.242	&1.6   &5-10        &601\\
     methanol	          &338       &18.9	       &748	    &0.295	&1.6  &6-10         &511\\
     ethanol              &352	     &18.6	       &750	    &0.420	      &1.6  &7-11           &355\\
     isopropanol        &356       &15.7	       &723	    &0.460 	&1.5  &9,10	         &283\\
     \hline
     \label{tab1}
\end{tabular*}
\end{table}
Figure \ref{pressure_oscillation}b shows the Fourier power spectrum of the pressure prior to the initiation of a star oscillation. There is a visible peak at $f_p\approx30$ Hz, as well as small peaks at higher frequencies. Figure \ref{pressure_oscillation}c shows data for the same drop later in time during a well-developed $n=4$ mode oscillation. Large variations in the pressure are visible, although the mean pressure remains unchanged. Figure \ref{pressure_oscillation}d shows the power spectrum during the star-shaped oscillation, indicating a sharp peak at $f_p\approx 28$ Hz, in addition to noise and harmonics at higher frequencies. The star-shaped oscillation frequency for this drop was $f_d\approx14$ Hz $\approx f_p/2$. 

\begin{figure}[!]
\begin{center}
\includegraphics[width=3.4 in]{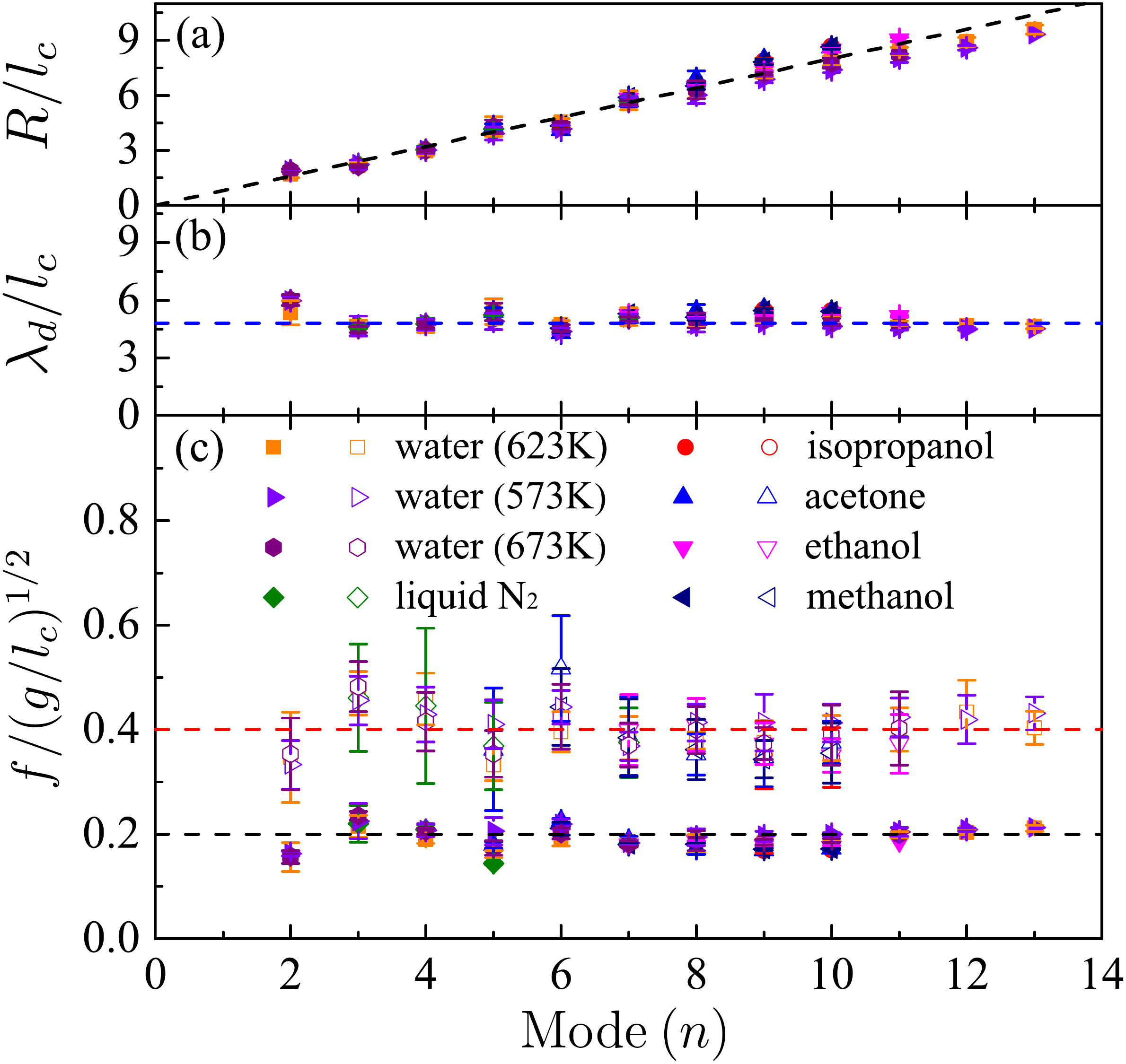}
\caption[]{Normalized radius (a), wavelength (b), and frequency (c) versus oscillation mode number $n$. Frequencies are plotted for both the star-shaped drop ($f_d$, closed symbols) and vapor layer pressure oscillations ($f_p$, open symbols). The error bars are described in the main text. The dashed line in (a) is the best linear fit to the data, whereas the dashed lines in (b) and (c) represent the mean values of the data.
\label{scaling}
} 
\end{center} 
\end{figure}

To gain further insight into the origin of the star-shaped oscillations, we used five other liquids in addition to water, as listed in Table \ref{tab1}. A summary of our results for the radius $R$, wavelength $\lambda_d$, drop frequency $f_d$, and characteristic pressure frequency $f_p$ is shown in Fig.\ \ref{scaling}. The error bars of $R$, $\lambda_d$, and $f_d$ come from the standard deviation of multiple measurements from different drops, whereas the error bar of $f_p$ is taken from the full width of the highest peak in the power spectrum at half the maximum value, as shown in Fig.\ \ref{pressure_oscillation}d. For all the drops, $f_d\approx f_p/2$, in agreement with a parametric forcing mechanism \cite{brunet2011star,shen2010parametrically}. The data collapses very well when scaled only by $l_c$, and the frequency scale $(g/l_c)^{1/2}$.  This collapse works for all the liquids used in the experiments, as well as water drops at different substrate temperatures, suggesting a purely hydrodynamic (non-thermal) mechanism for generating star-shaped oscillations. This is in agreement with recent experiments studying star-shaped oscillations of drops levitated by an air flow from below \cite{bouwhuis2013oscillating}. 

\begin{figure}[!]
\begin{center}
\includegraphics[width=3.4 in]{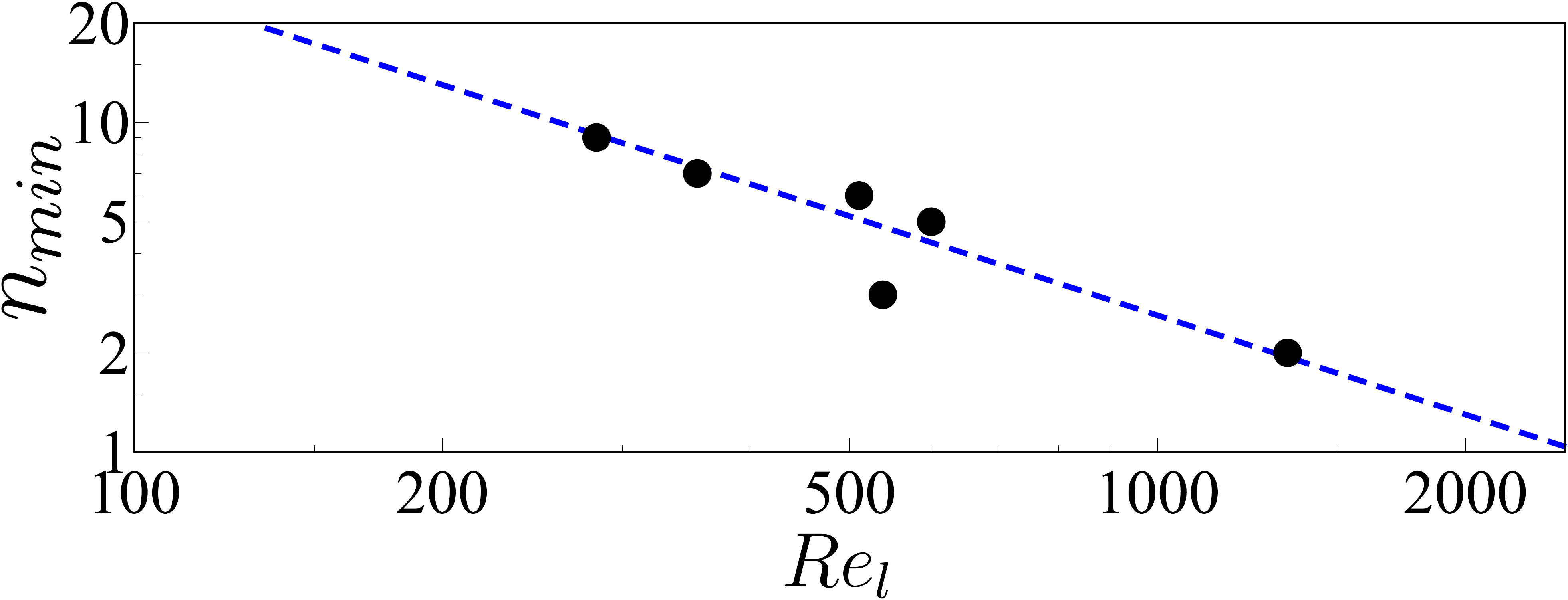}
\caption[]{Scaling behavior for the minimum observed star mode $n_{min}$ with respect to $Re_l$ of different liquids (Table \ref{tab1}). The blue dashed line represents $2600 Re_l^{-1}$.
\label{threshold}
} 
\end{center} 
\end{figure}

Since $\lambda_d$ is approximately constant, modes with large $n$ could only be observed in large drops. Occasionally we observed the same mode number at smaller values of $R$, but this only happened for $n=2-4$ and was difficult to replicate. In addition, we found that the number of observed modes was sensitive to the viscosity of the liquid. Assuming a characteristic length, $l_c$, and time scale, $\sqrt{l_c^3 \rho_l/\gamma}$, we express the Reynolds number of each liquid, $Re_l\equiv (l_c \gamma \rho_l)^{1/2}/\eta_l$, as shown in Table \ref{tab1}. We find that the minimum observed mode number, $n_{min}$, is inversely proportional to $Re_l$, as shown in Fig.\ \ref{threshold}. This suggests that liquid viscosity damps the oscillations of smaller drops and thus sets the threshold mode for self-sustained star oscillations. The largest observable value of $n$ is ultimately determined by the size of the experimental apparatus, except in the case of liquid N$_{2}$. In this case, the evaporation rate is very high since the susbstrate temperature is much higher than the boiling point. Even though the aluminum surface is curved, for large drops the resulting vapor layer is thick enough to develop bubbles due to the Rayleigh-Taylor instability \cite{trinh2014curvature}.

One surprising result drawn from the experimental data is that only a single dominant wavelength (and thus frequency) exists for all modes, and only depends on the capillary length of the liquid. There are small oscillations of both $\lambda_d$ and $f_d$ about their mean values (dashed lines in Figs.\ \ref{scaling}b and\ \ref{scaling}c). These oscillations are strongest for small $n$, and are consistent among the different liquids and substrate temperatures. This phenomenon may be caused by nonlinear effects, e.g., the dependence of the oscillation frequency on the amplitude is stronger for smaller modes \cite{becker1991experimental,smith2010modulation}. Nevertheless, the data suggests a very robust mechanism for selecting either the frequency or wavelength of the modes. 

The relationship between $f_d$ and $f_p$ can be understood from the quasi-2D dispersion relation for large, puddle-shaped drops, where $f_d\propto R^{-3/2}$ \cite{bouwhuis2013oscillating,yoshiyasu1996self,ma2015many}. 
Assuming that the radius of the puddle varies sinusoidally with time, then modes of the star oscillations follow an equation similar to the Mathieu equation, and will be excited when $f_d\approx f_p/2$ \cite{yoshiyasu1996self,brunet2011star}. 
In the case of Leidenfrost drops, there is no obvious frequency or wavelength selection mechanism generated by the flow and evaporation of vapor beneath the drop. It is possible that a ``breathing mode'' of the drop would cause the radius to vary with time, however, recent measurements of the breathing mode in both low and high-viscosity levitated drops show that the frequency rapidly decreases with $R$ \cite{caswell2014dynamics,bouwhuis2013oscillating,Maleidenfrost}, in contrast to the data shown in Fig.\ \ref{scaling}c. 

\begin{figure}[!]
\begin{center}
\includegraphics[width=3.4 in]{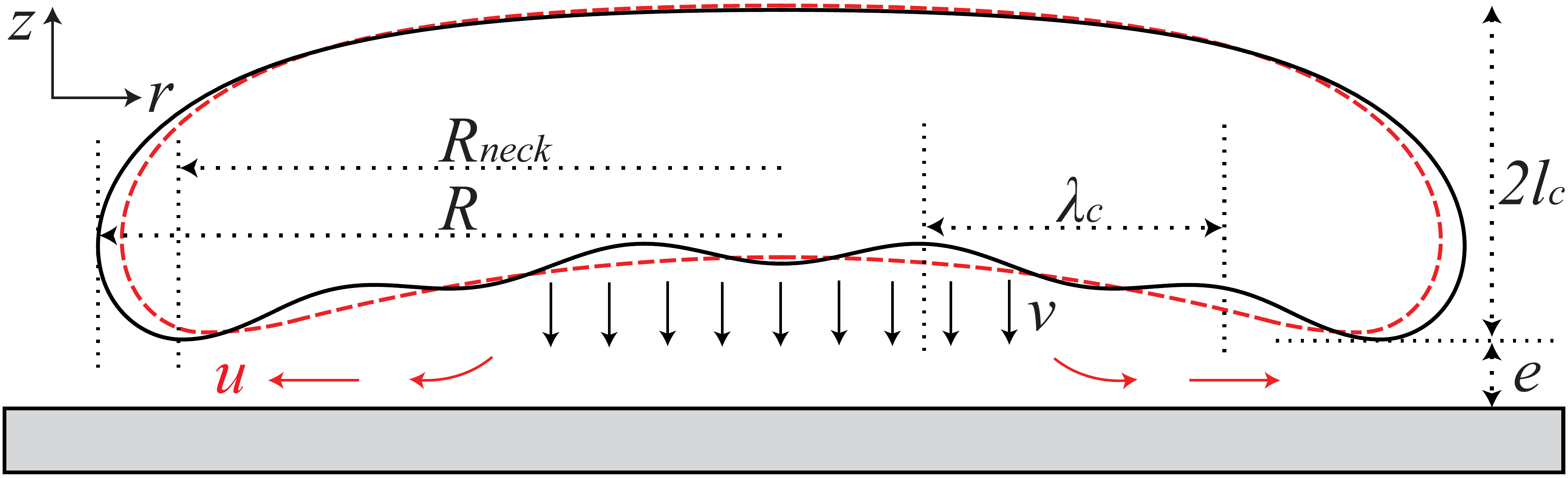}
\caption[]{Cartoon cross-section of a Leidenfrost drop before (red dashed curve) and after (black curve) the excitation of capillary waves under the drop. Symbols are defined in the text.
\label{hydrodynamics}
} 
\end{center} 
\end{figure}

The pressure variations beneath the drop are determined by the shape of the liquid-vapor interface and the flow in the vapor layer. A simple model of the flow beneath the drop is shown in Fig.\ \ref{hydrodynamics}, where $v$ is the velocity of the gas at the liquid surface, $e$ is the mean thickness of the vapor layer, and $u$ is the radial velocity of the gas near $r=R$. Following the model in Biance \textit{et al.} \cite{biance2003leidenfrost} by assuming incompressibility and lubrication flow in the gas, $e$, $v$, and $u$ can be expressed as
\begin{eqnarray}
\label{ueq}
u=\left( \frac{\rho_l g\kappa l_c}{3\rho_v\eta_vL} \right)^{1/2}\Delta T^{1/2},\\
\label{veq}
v=\left[ \frac{4l_c\rho_l g}{3\eta_v}\left( \frac{\kappa \Delta T}{L \rho_v} \right)^{3} \right]^{1/4}R^{-1/2},\\
\label{eeq}
e=\left( \frac{3\kappa \Delta T \eta_v}{4L\rho_l\rho_vgl_c} \right)^{1/4}R^{1/2},
\end{eqnarray}  
where $\Delta T=T_s-T_b$, $L$ is the latent heat of evaporation, and $\eta_v$, $\rho_v$, and $\kappa$ are the mean values of the dynamic viscosity, density, and thermal conductivity in the vapor layer. Although the approximation assumes a steady-state, linear temperature profile in the vapor layer, this is valid since the typical thermal diffusion time across the vapor layer is $\approx$ 1 ms for water vapor near the boiling point. 

The Reynolds number in the vapor layer can be expressed as $Re_v\equiv v e \rho_v/\eta_v$ \cite{lister2008shape,duchemin2005static}. After plugging in the expressions for $v$ and $e$, $Re_v=\Delta T\kappa/L\eta_v$, which is independent of the drop radius. For typical values of the properties in a water vapor layer,  $Re_v\approx0.2$. Thus both viscous and inertial forces are relevant. However, a simple estimate of the Bernoulli pressure, $\rho_v u^2/2\approx 1$ Pa, suggests that inertia is not responsible for the pressure variations observed in Fig.\ \ref{pressure_oscillation}. 

On the other hand, the viscous lubrication pressure scales as $\eta_v v R^2/e^3$, so that variations in pressure are related to small thickness changes: $\Delta p$ = ($3  \eta_v v R^2/e^4$)$\Delta e$. For the water drop shown in Fig.\ \ref{pressure_oscillation}d, where $\Delta p\approx 10$ Pa, we would expect thickness variations of order $\Delta e\approx 15$ $\mu$m. These variations could be produced by a vertical motion of the center of mass of the drop, although, the frequency of this oscillation should decrease for larger drops, in contrast to the data in Fig.\ \ref{scaling}. Thus it is likely that the pressure oscillations occur due to local changes in vapor thickness. 

\begin{figure}[!]
\begin{center}
\includegraphics[width=3.4 in]{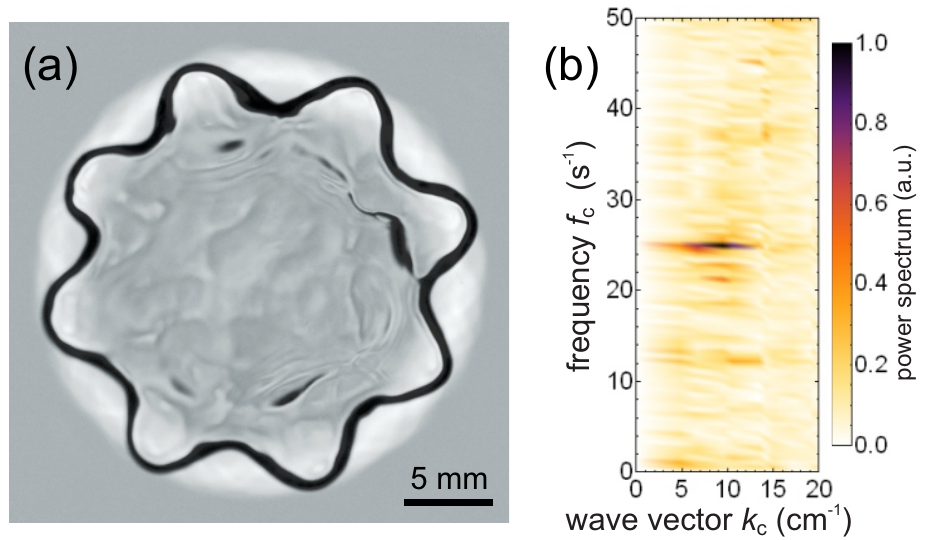}
\caption[]{(a) Capillary waves imaged beneath an $n=8$ mode acetone drop. The image has been enhanced for visibility. (b) Corresponding power spectrum associated with the acetone Leidenfrost drop. There is a sharp peak at $f_c\approx$ 26 Hz which covers a broad range of wave vectors.  
\label{capillary_waves}
} 
\end{center} 
\end{figure}

We hypothesize that the pressure variations are caused by capillary waves of a characteristic wavelength (frequency), $\lambda_c$ ($f_c$), which travel from the center of the drop to the edge. When they reach the edge, the waves will cause the radius to vary with time, as shown schematically in Fig.\ \ref{hydrodynamics}. To support this hypothesis, we have measured the spectrum of capillary waves by imaging the vapor layer under a star-shaped Leidenfrost drop on a curved, fused silica surface. Variations in the thickness of the vapor layer deflect the collimated light used for illumination. A typical image under an acetone drop is shown in Fig.\ \ref{capillary_waves}a, and the corresponding power spectrum is shown in Fig.\ \ref{capillary_waves}b. The spectrum shows a large peak at $f_c\approx$ 26 Hz which is spread over a range of wave numbers located near $k_c\approx$ 10 cm$^{-1}$. 

The exact selection mechanism for the dominant wavelength $\lambda_c$ beneath the drop is the result of interactions between the lubricating flow of the evaporating vapor, and the deformable liquid interface. Capillary waves with small wavelengths are known to be unstable when a liquid interface is driven by a strong shear stress \cite{Miles1957,Zhang1995,Zeisel2008,Paquier2015}. For water waves, the strength of the shear is often measured by the friction velocity, $u_\star=\sqrt{\tau/\rho_v}$, where $\tau$ is the shear stress at the interface. Assuming a parabolic-flow profile in the vapor layer with mean velocity $u$ near the edge of the drop (Fig.\ \ref{hydrodynamics}), the maximum shear stress at the liquid-vapor interface is $\tau=6\eta_v u/e$. Plugging in the expressions for $u$ and $e$ from Eqs.\ \ref{ueq} and \ref{eeq}, and using typical values for the properties of water vapor, we estimate that $u_\star\approx1-2$ m/s for a Leidenfrost drop. This friction velocity is quite strong, and can easily lead to the unstable growth of modes with wavelengths of order a few millimeters \cite{Zeisel2008,Zhang1995}. The threshold for the onset of instability, as shown in Fig. \ref{threshold}, should also be related to the forcing and velocity in the vapor layer \cite{bouwhuis2013oscillating}. This remains an open question for future studies.

We can estimate for the dominant wave vector using the zeroth-order dispersion relation for capillary waves under the drop. We assume a constant pressure in the vapor layer, and that the top surface of the drop is nearly flat:
\begin{equation} 
f_c=\frac{1}{2\pi}\sqrt{\left(-gk_c+\frac{\gamma k_c^3}{\rho_l} \right)\tanh\left( k_ch \right)}.
\label{cwaves}
\end{equation}
Here $k_c$ = 2$\pi$/$\lambda_c$, and $h\approx2 l_c$ is the thickness of the drop. This is the well-known dispersion relation for capillary-gravity waves, except the denser fluid rests above the lighter gas. Since we are not considering the restoring force associated with the lubrication pressure, our estimate serves as an upper bound for $k_c$. For a large Leidenfrost water drop, the measured frequency of pressure variations is $f_p\approx26$ Hz. Using Eq. \ref{cwaves}, and assuming $f_p=f_c$, we arrive at $\lambda_c \approx 3.03l_c$. For the acetone drop shown in Fig.\ \ref{capillary_waves}a, this corresponds to $k_c\approx$ 12.9 cm$^{-1}$, which is slightly larger than the position of the peak in Fig.\ \ref{capillary_waves}b. 

This result is also in agreement with the minimum value of $R$ required to observe an $n=2$ mode. The drop radius we measured is the maximum radius, $R$, instead of $R_{neck}$, which is the radius at which the drop comes closest to the surface (Fig.\ \ref{hydrodynamics}). The minimum size necessary to fit one wavelength beneath the drop is $R_{neck}=\lambda_c/2$. The relationship between $R$ and $R_{neck}$ for large Leidenfrost drops is $R\approx R_{neck}+0.53l_c$ \cite{burton2012geometry,snoeijer2009maximum}. This means that the minimum value of $R$ required to observe a star-shaped oscillation is $R\approx2.05l_c$, in agreement with data shown in Fig.\ \ref{scaling}a.

In summary, we experimentally investigated the self-sustained, star-shaped oscillations of Leidenfrost drops of different liquids on curved surfaces. The star oscillation wavelength and frequency are nearly independent of the drop radius, mode number, and substrate temperature, and only depend on the capillary length of the liquid, indicating a purely hydrodynamic (non-thermal) origin for the oscillations. We conclude that capillary waves beneath the drop occur with a characteristic wavelength and are the cause of the pressure oscillations, which parametrically induce the star-shaped oscillations. These results may enhance our understanding of thin, supporting gas films in contact with a liquid interface and a solid surface, a scenario which also occurs during drop impact and gas entrainment.

We gratefully acknowledge financial support from National Science Foundation, DMR-1455086.

\bibliography{star}

\end{document}